\documentclass[aps,prd,twocolumn,superscriptaddress,nofootinbib,a4paper,nobibnotes,longbibliography]{revtex4-1}  

\usepackage{graphicx}
\usepackage{dcolumn}
\usepackage{bm}
\usepackage{latexsym}
\usepackage{amsfonts}
\usepackage{subcaption}
\usepackage{amssymb}
\usepackage{amsmath}
\usepackage{bbold}
\usepackage{color}
\usepackage[utf8]{inputenc} 
\usepackage{xspace}

\def\beq{\begin{equation}}
\def\eeq{\end{equation}}
\def\bea{\begin{eqnarray}}
\def\eea{\end{eqnarray}}

\begin{document}
\title{Ho\v{r}ava Gravity after GW170817}
\author{A. Emir G\"umr\"uk\c{c}\"uo\u{g}lu}
\affiliation{Institute of Cosmology and Gravitation, University of Portsmouth, Portsmouth PO1 3FX, UK}
\author{Mehdi Saravani}
\affiliation{School of Mathematical Sciences, University of Nottingham, University Park, Nottingham, NG7 2RD, UK}
\author{Thomas P. Sotiriou}
\affiliation{School of Mathematical Sciences, University of Nottingham, University Park, Nottingham, NG7 2RD, UK}
\affiliation{School of Physics and Astronomy, University of Nottingham, University Park, Nottingham, NG7 2RD, UK}

\begin{abstract}
The recent detection of gravitational wave GW170817 has placed a severe bound on the deviation of the speed of gravitational waves from the speed of light. We explore the consequences of this detection for Ho\v{r}ava gravity. 
\end{abstract}

\maketitle

\section{Introduction}

Ho\v{r}ava gravity \cite{Horava:2009uw} has been proposed as an ultraviolet (UV) complete theory for the gravitational interaction. The improved behaviour at high energies is due to the presence of higher-order terms in spatial derivatives and this requires  violations  of local Lorentz invariance. 
The  action of Ho\v{r}ava gravity is \cite{Horava:2009uw,Blas:2009qj}
\begin{align}
S = \frac{M_{p}^2}{2}\int N\,dt\,\sqrt{h}d^3x
&\bigg(K_{ij}K^{ij}-\lambda\,K^2+\xi\,^{(3)}R+\eta\,a_ia^i
\nonumber\\
&
 +\frac{1}{M_\star^2}{\cal L}_4+\frac{1}{M_\star^4}{\cal L}_6\bigg).\,
\label{eq:fullaction}
 \end{align}
$N$, $N^i$, and $g_{ij}$ are the lapse function, the shift and the induced metric on a given spacetime foliation by spacelike hypersurfaces, while $K_{ij}$ is their extrinstic curvature. ${\cal L}_4$ and ${\cal L}_6$ contain terms that are fourth-order and sixth-order in spatial derivatives, respectively. Hence, they contribute  fourth and sixth powers of momenta in the dispersion relations. These contributions are suppressed at momenta below some scale $M_*$, where the theory becomes effectively an infrared (IR) modification of General Relativity (GR), but dominate in the UV and are expected to render interactions renormalizable. The action is not invariant under the full group of diffeomorphisms, but only under diffeomorphisms that respect the foliation. This underscores that the foliation is preferred. 

The first line of eq.~\eqref{eq:fullaction} can be thought of as the IR limit of Ho\v rava gravity. Elevating the space-time symmetry to full diffeomorphisms by introducing a St\"uckelberg field $\phi$, one can write the IR action as \cite{Blas:2010hb}
\begin{align}
S_{IR}=\frac{M^2_\textrm{\AE{}}}{2}\int & \sqrt{-g}\,d^4x\bigg[~^{(4)}R+\alpha\,u^\mu u^\nu \nabla_\mu u_\alpha\,\nabla_\nu u^\alpha \nonumber\\
&-\beta\,\nabla_\mu u_\nu\,\nabla^\nu u^\mu -\gamma (\nabla_\mu u^\mu)^2 \bigg]\,,
\label{eq:khronometric}
\end{align}
where we defined $u_\mu \equiv -\nabla_\mu \phi/\sqrt{-\nabla^\nu \phi\nabla_\nu \phi}$. This action coincides with the first line of action \eqref{eq:fullaction} after the partial gauge fixing $\phi=t$ and with the following correspondence of  parameters through
\begin{eqnarray}
\label{corres}
\lambda &= \dfrac{1+\gamma}{1-\beta}\,,\qquad
\eta &= \dfrac{\alpha}{1-\beta}\,,\nonumber\\
\xi&= \dfrac{1}{1-\beta}\,,\qquad
M_{p}^2 &= (1-\beta)M^2_\textrm{\AE{}}\,.
\end{eqnarray}
Moreover, the covariant formulation is equivalent to a restricted version of the Einstein-\AE{}ther theory \cite{Jacobson:2000xp}, in which the \ae{}ther is forced to be hypersurface-orthogonal at the level of the action ({\em i.e.}~before the variation) \cite{Jacobson:2010mx}. The correspondence of parameters is given in  Appendix \ref{sec:aetherapp}.

Ho\v rava gravity propagates two tensor and one scalar  polarization. All three polarizations satisfy higher-order dispersion relations, as mentioned earlier. Their speeds in the infrared limit are
\begin{equation}
c_T^2 = \frac{1}{1-\beta}\,,\qquad
c_S^2 = \frac{(2-\alpha)(\gamma+\beta)}{\alpha(1-\beta)(2+3\,\gamma+\beta)}\,.
\label{speeds}
\end{equation}
The recent detection of a binary neutron star merger with coincident gamma ray emission has introduced remarkably strong constraints on $c_T$ \cite{Monitor:2017mdv}. The purpose of this brief note is to discuss the implications of this constraint for Ho\v rava gravity and to clarify how this constraint can be effectively combined with existing ones. It has recently been pointed out in Ref.~\cite{Sotiriou:2017obf} that the speed of the scalar polarisation is almost unconstrained and our results highlight that this feature persists even after  GW170817. 

It is worth pointing out that our focus is on infrared viability. Hence, the higher order terms contained in ${\cal L}_4$ and ${\cal L}_6$ in action \eqref{eq:fullaction} will not be relevant and we will not attempt to give them explicitly. We will, however, discuss constraints on $M_\star$ and mention how theoretical considerations related to the UV properties of the theory can restrict $M_\star$. We will also not consider any version of Ho\v rava gravity where additional restrictions of the action are considered in order to reduce the numbers of independent couplings, {\em e.g.}~\cite{Sotiriou:2009gy,Weinfurtner:2010hz,Horava:2010zj,Vernieri:2011aa,Vernieri:2012ms}. Even though some of these restricted models have been shown to have interesting properties --- the so called `projectable' theory \cite{Sotiriou:2009gy}, for instance, has been shown to be renormalizable beyond power-counting \cite{Barvinsky:2015kil}\footnote{Projectable theory in 2+1 dimensions \cite{Sotiriou:2011dr} has actually been shown to be asymptotically free \cite{Barvinsky:2017kob}.} --- they also tend to suffer from infrared viability issues \cite{Sotiriou:2009bx,Charmousis:2009tc,Blas:2009yd,Koyama:2009hc,Papazoglou:2009fj,Kimpton:2010xi,Mukohyama:2010xz}.

\section{Direct constraints and bounds}\label{section_constraint}
In this section we list  all of the available constraints in terms of the parameters $(\alpha,\beta,\gamma)$ of action \eqref{eq:khronometric}.

\begin{enumerate}

\item \underline{Unitarity.} The kinetic term for the scalar mode should have the same sign as the kinetic term of the tensor modes in order for the Hamiltonian to be bounded for linearised perturbations around flat space. This yields \cite{Blas:2009qj}:
\begin{equation}
\frac{2+3\,\gamma+\beta}{\gamma+\beta} >0\,.
\end{equation}
\item \underline{Perturbative stability.} The coefficients of the gradient terms should have the right sign for stability \cite{Blas:2009qj}:
\begin{equation}
 0<\alpha<2\,,\qquad
 \beta<1\,.
\end{equation}
This condition, along with the previous one ensures that $c_T^2$ and $c_S^2$ are always positive.
\item \underline{BBN.} 
Cosmology provides further constraints. On a cosmological background, the effective gravitational constant that appears in the Friedmann equation is \cite{Blas:2009qj}
\begin{equation}
G_C = \frac{1}{4\,\pi M_{\textrm{\AE{}}}^2(2+3\gamma+\beta)}\,,
\end{equation}
while from the Newtonian limit, one can infer \cite{Jacobson:2008aj}
\begin{equation}
G_N = \frac{1}{8\,\pi\,M_{\textrm{\AE{}}}^2(1-\alpha/2)}\,.
\end{equation}
This effective gravitational constant affects the expansion rate during Big Bang Nucleosynthesis (BBN) with respect to the standard one. As a result the primordial helium abundance is modified by \cite{Chen:2000xxa,Carroll:2004ai}
\begin{equation}
\Delta Y_p = 0.08 \left(\frac{G_C}{G_N}-1\right)\,.
\end{equation}
Using the current bound $|\Delta Y_p| < 0.01$ ($99.7\%$ C.L.) \cite{Izotov:2014fga,Aver:2015iza,Patrignani:2016xqp} we obtain the following constraint
\begin{equation}
\left\vert \frac{\alpha+3\,\gamma+\beta}{2+3\,\gamma+\beta}\right\vert <\frac{1}{8}\,.
\end{equation}
\item \underline{Vacuum Cherenkov bounds.} Photons could decay into  spin-2 or spin-0 modes in vacuum when Lorentz symmetry is violated. Cosmic rays provide a lower bound on the speed of gravitational polarizations \cite{Moore:2001bv}. Specific constraints for Einstein-\AE{}ther theory have been derived in Ref.~\cite{Elliott:2005va} and they exclude subluminal propagation to very high precision.\footnote{There is no compelling reason to {\em a priori} exclude superluminal propagation, unlike what seems to be suggested in Ref.~\cite{Elliott:2005va}. Hence, vacuum Cherenkov constraints are one-sided.} There is no detailed calculation for Ho\v rava gravity, or a quantitative translation of the Einstein-\AE{}ther results (given the similarity of the theories). However, the conservative expectation is that subluminal propagation is excluded to very high accuracy for both tensor and scalar polarisations.

The absence of gravitational Cherenkov radiation can, in principle, give a bound on $M_*$, the scale that suppressed the higher-order corrections to the dispersion relation of gravitational waves \cite{Kiyota:2015dla,Yunes:2016jcc}. However, to obtain a noteworthy bound one needs to assume that  the coefficient of $p^4$ term (where $p$ is momentum) in the dispersion relation is negative. Moreover, such a constraint  would only be trustworthy if $M_*$ is much bigger than the cosmic ray energies ($\sim 10^{11}$ GeV). Otherwise, one would need to include the $p^6$ term in the analysis as well, which is expected to have a positive sign. We will not consider this type of bound below.

 \item \underline{ppN constraints.} The two parametrized post-Newtonian (ppN) parameters which quantify preferred-frame effects are constrained by \cite{Will:2005va}
 \bea
&&| \alpha_1 | < 10^{-4}\,,\notag\\
&&|\alpha_2| < 4\times 10^{-7}\,.
 \eea
Using the weak field expressions for Ho\v{r}ava gravity \cite{Blas:2010hb,Blas:2011zd}, these constraints translate into
\begin{align}
\left \vert \frac{4\,(\alpha-2\,\beta)}{1-\beta} \right\vert &\lesssim 10^{-4}\,,\nonumber\\
\left \vert \left(\frac{\alpha-2\,\beta}{2-\alpha}\right)\left(1-\frac{(\alpha-2\,\beta)(1+\beta+2\,\gamma)}{(1-\beta)(\beta+\gamma)}\right)\right\vert &\lesssim 10^{-7}\,.
\end{align}
\item \underline{Binary pulsars.} The presence of a scalar polarisation can lead to dipolar emission and this would affect the dynamics of binary pulsars. The corresponding bounds are discussed in details in Ref.~\cite{Yagi:2013ava}.  Figs.~1b and 8 of Ref.~\cite{Yagi:2013ava} present the constraints on the $\alpha=2\,\beta$ plane. We will not reproduce these figures here. On this plane the ppN parameters $\alpha_1$ and $\alpha_2$ vanish but, as will be discussed below, considering this plane is no longer well motivated. 
\item \underline{Black holes.} The  structure of isolated black holes can, in principle, provide constraints \cite{Barausse:2011pu,Barausse:2012ny,Barausse:2012qh,Barausse:2013nwa}. However, such constraints are significantly weaker than the binary pulsar constraints (at least on the $\alpha=2\,\beta$ plane).
\item \underline{Gravitational Waves.} The observation of the binary neutron star merger GW170817 with coincident gamma ray emission \cite{Monitor:2017mdv} yields 
\begin{equation}
\label{gwcon}
-3\times 10^{-15} \leq c_T-1\leq 7\times 10^{-16}\,,
\end{equation}
 which implies that 
\begin{equation}
|\beta| \lesssim 10^{-15}\,.
\end{equation}
Gravitational wave observations also provide a very mild lower bound on $M_\star$ of the order of $eV$ \cite{Abbott:2017vtc,Sotiriou:2017obf}. This bound comes from considering the effect of the $p^4$ term in the dispersion relation (where $p$ is momentum) that is suppressed by $M_\star^2$. 
\end{enumerate}

\begin{figure}[t!]
	\begin{subfigure}{0.45\textwidth}
	\includegraphics[width=\textwidth]{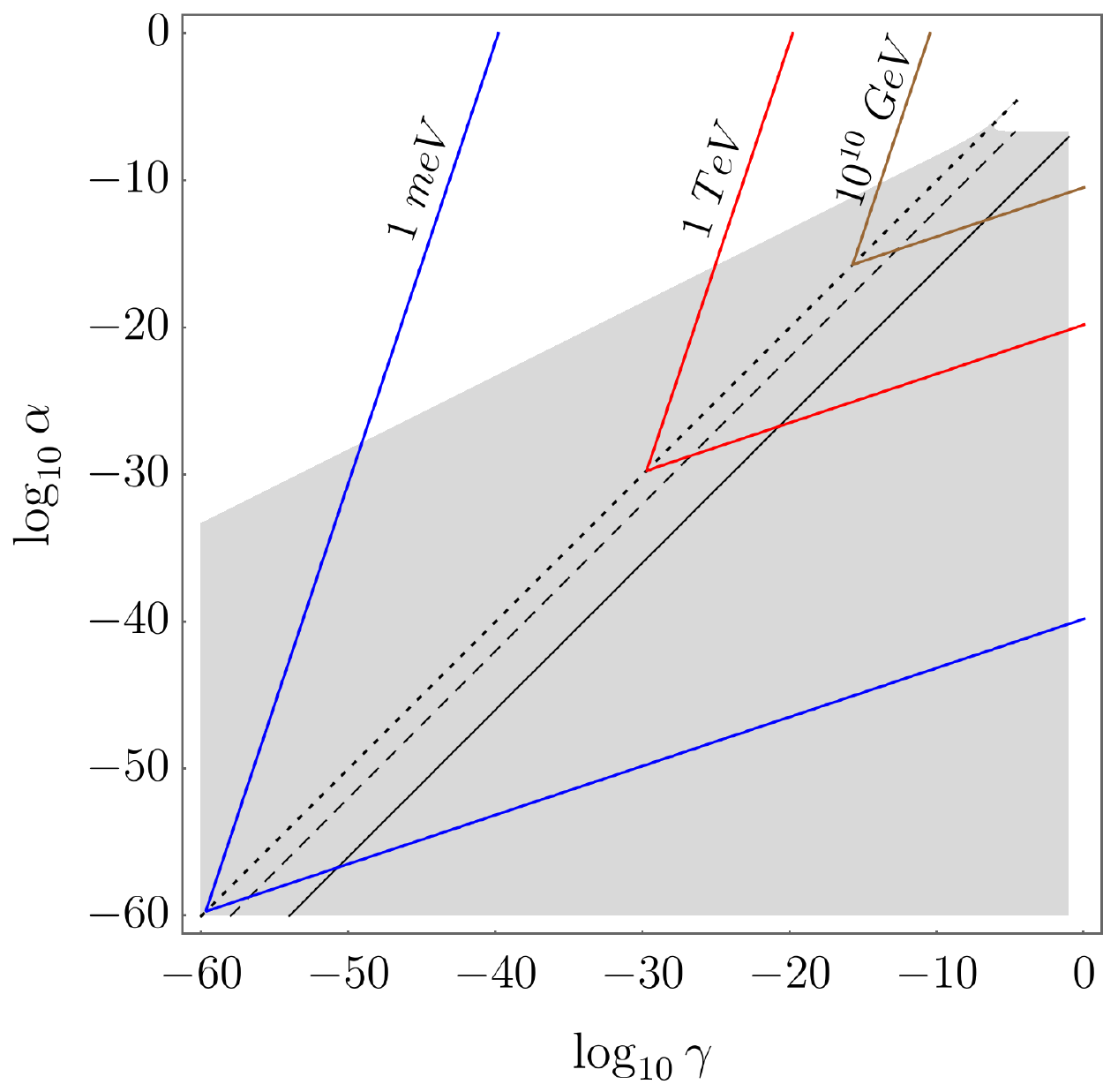}
	\end{subfigure}
	\begin{subfigure}{0.45\textwidth}
	\includegraphics[width=\textwidth]{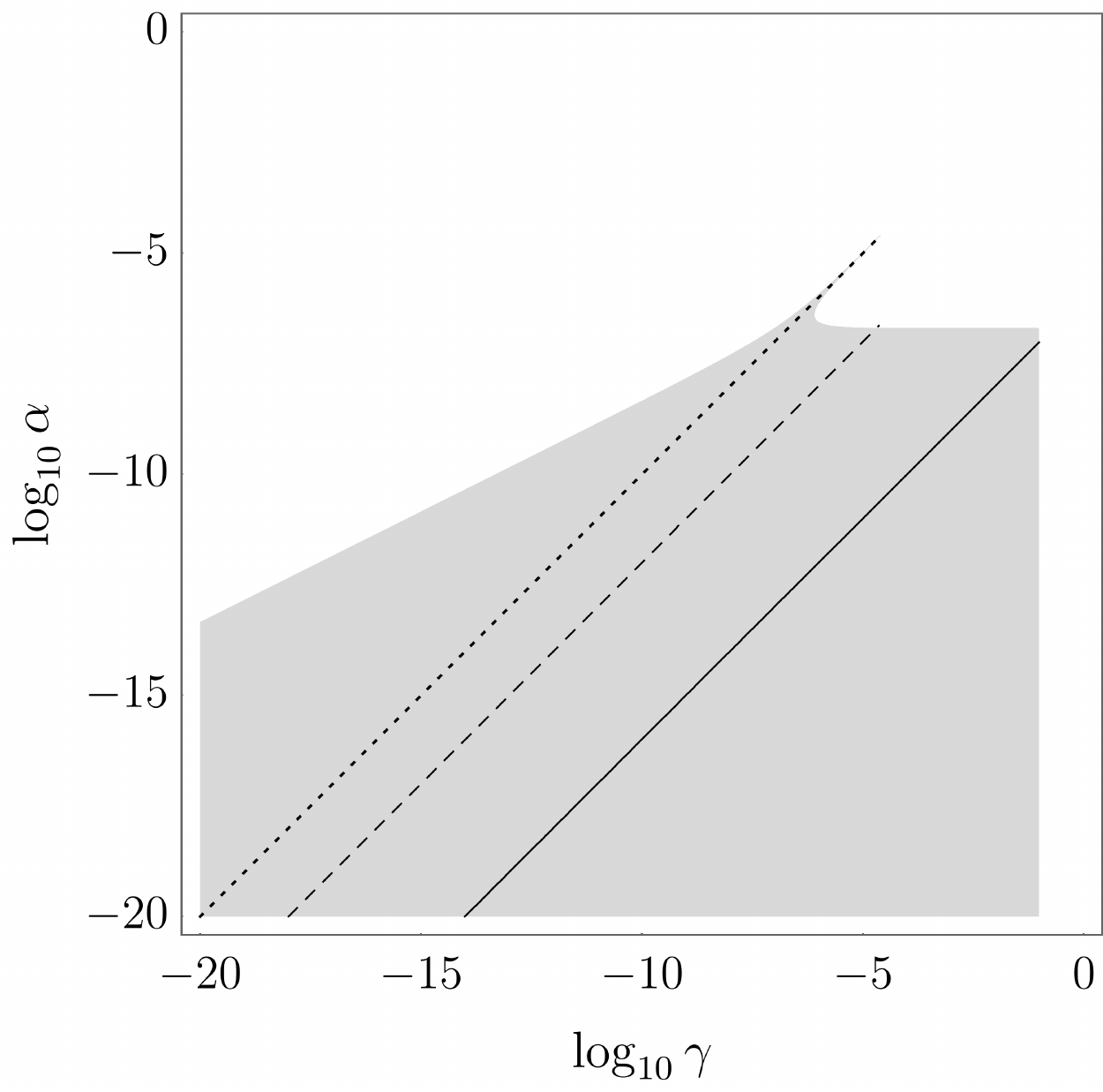}
	\end{subfigure}
\caption{The grey area shows the region of parameter space which is compatible with the constraints (Sec.  \eqref{section_constraint}) for $\beta=0$. The dotted, dashed and solid lines  correspond to $c_S=1$, $c_S=10$ and $c_S=1000$, respectively. The blue, red and brown lines show $M_{SC}=1$ meV, $M_{SC}=1$ TeV and $M_{SC}=10^{10}$ GeV. The lower panel focuses on the region $\alpha\,,\gamma>10^{-20}$.}
\label{fig:parameter_space_zero}
\end{figure}

\begin{figure}[t!]
	\begin{subfigure}{0.45\textwidth}
	\includegraphics[width=\textwidth]{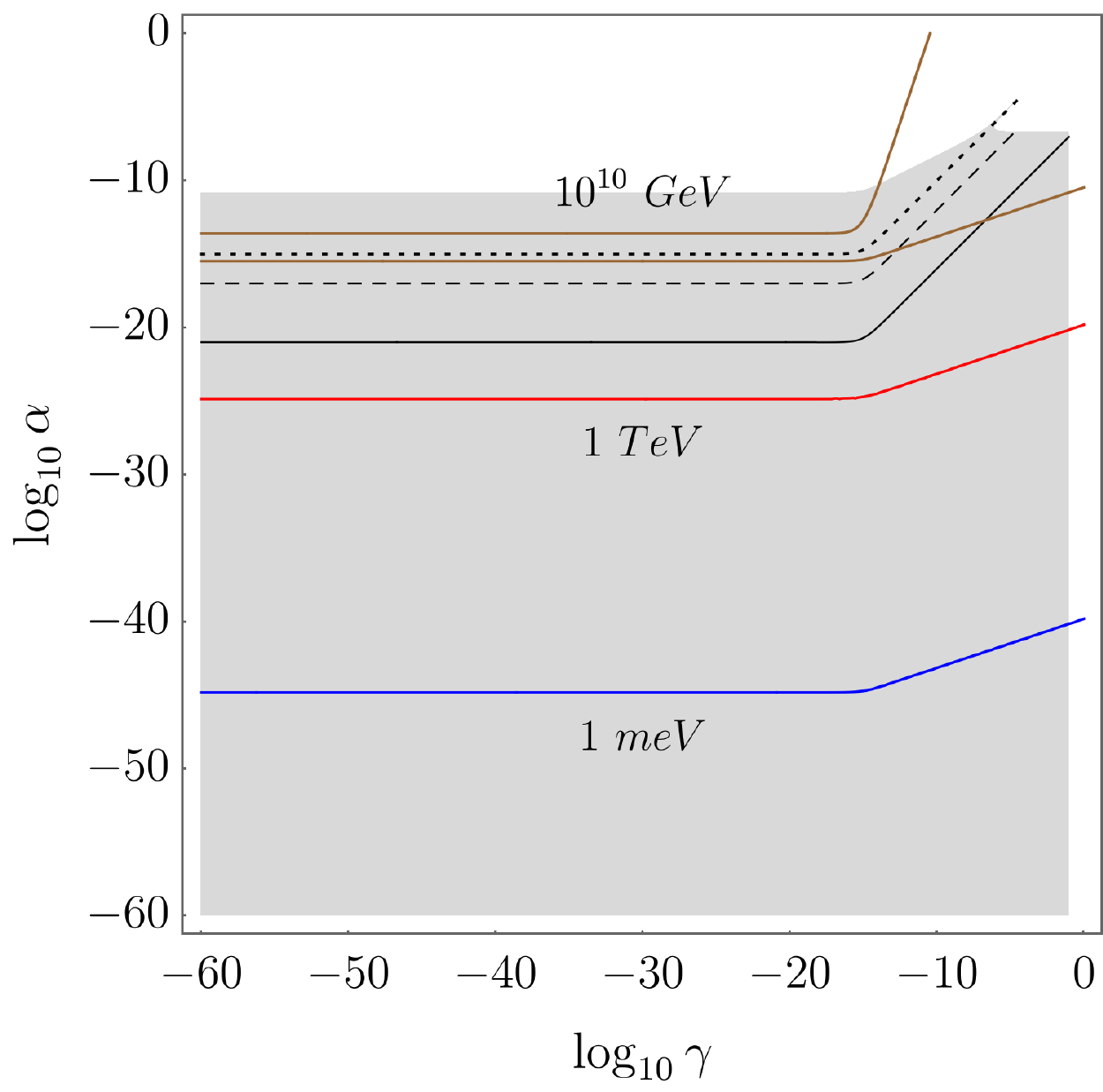}
	\end{subfigure}
	\begin{subfigure}{0.45\textwidth}
	\includegraphics[width=\textwidth]{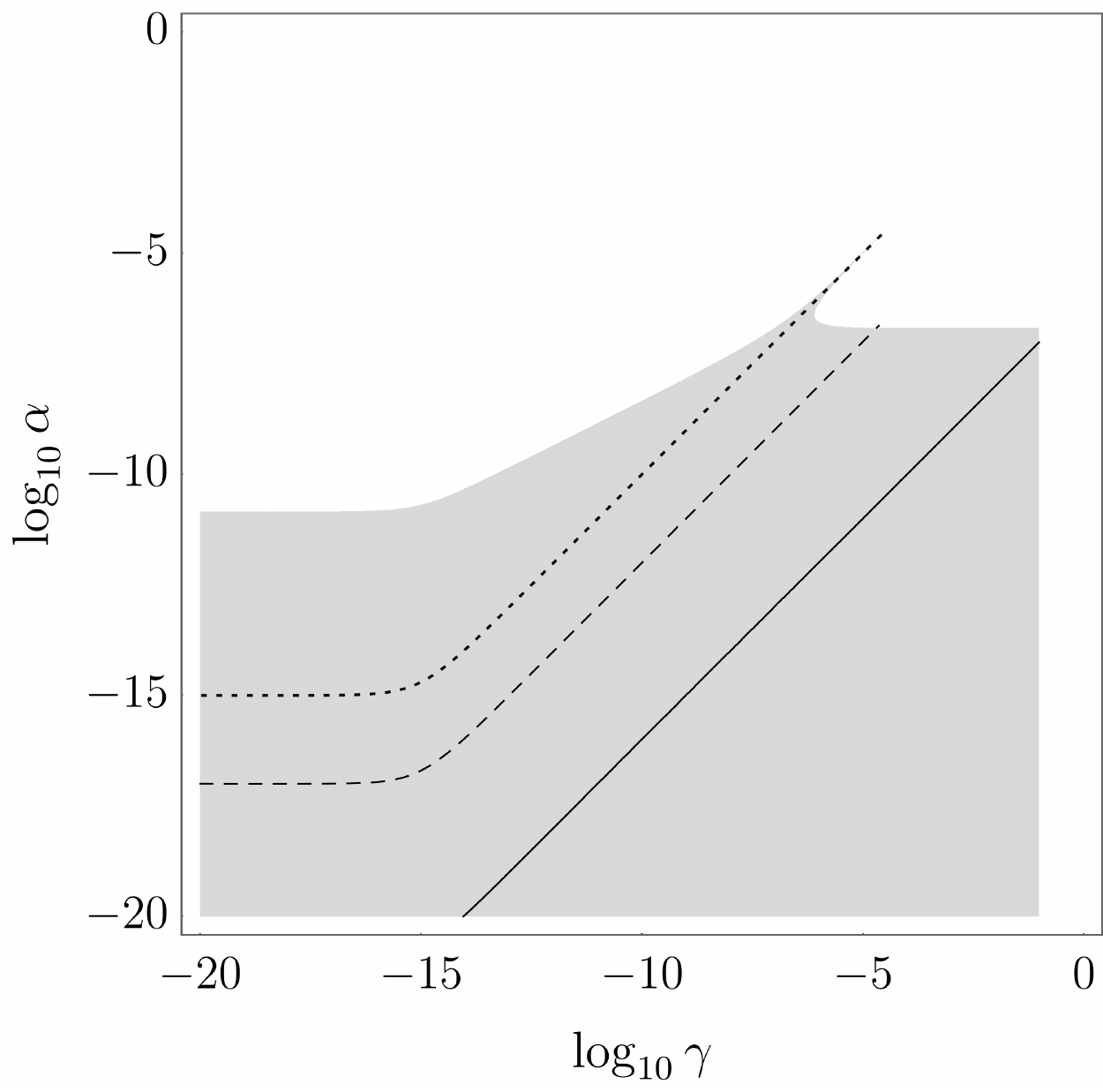}
	\end{subfigure}
\caption{Same as Fig.~\ref{fig:parameter_space_zero} but for $\beta=10^{-15}$.}
\label{fig:parameter_space_positive}
\end{figure}

\begin{figure}[t!]
	\begin{subfigure}{0.45\textwidth}
	\includegraphics[width=\textwidth]{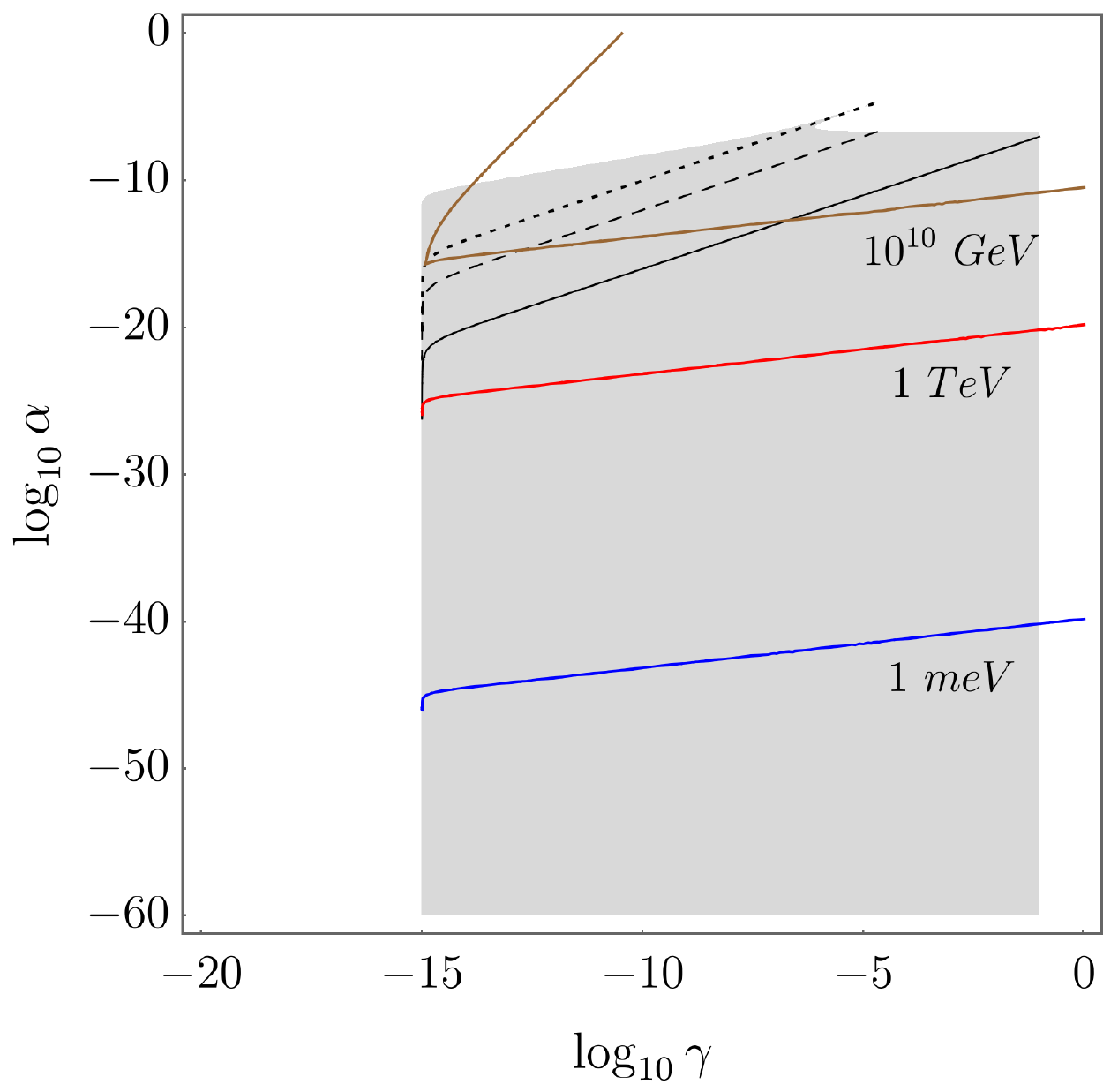}
	\end{subfigure}
	\begin{subfigure}{0.45\textwidth}
	\includegraphics[width=\textwidth]{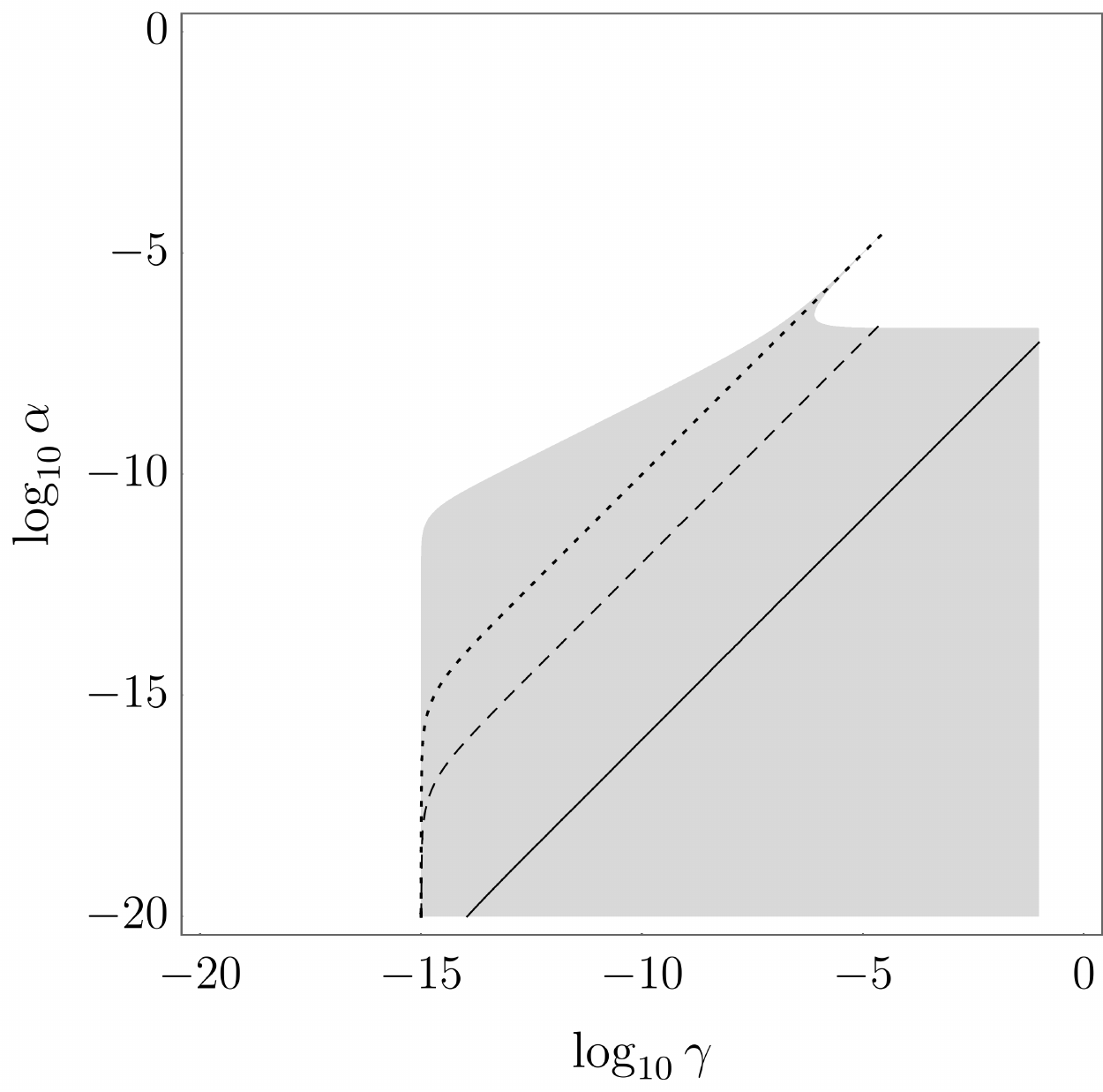}
	\end{subfigure}
\caption{Same as Fig.~\ref{fig:parameter_space_zero} but for $\beta=-10^{-15}$.}
\label{fig:parameter_space_negative}
\end{figure}

\section{Theoretical considerations and indirect constraints}
\label{sec:dec}

The direct observational bounds obtained in the previous Section affect mainly the parameters of the IR effective theory. The parameters that become relevant in the UV are very weakly affected. In particular, gravitational waves provide only a very weak bound of $M_\star$, as mentioned above. In principle, laboratory test of gravity at small length scales can place a direct lower bound on $M_\star$ (as an energy scale), but current precision would place this bound in the $meV$ range \cite{Adelberger:2009zz}, so it would be also particularly weak. 

 Lorentz violations in the standard model are much more tightly constrained than in gravity. Indeed, if one were to assume that $M_\star$ is a universal Lorentz violation scale for gravity and matter alike, then observations of the synchrotron radiation from the Crab nebula would require $M_\star >2\times 10^{16} ~GeV$ \cite{Liberati:2012jf}. However, if there is a mechanism to suppress the percolation of Lorentz symmetry breaking from gravity to matter, then there is no reason to believe that $M_\star$ is a universal Lorentz violation scale. Moreover, such a mechanism seems to be necessary to keep Lorentz violation at bay already for lower mass dimension operators in the standard model \cite{Iengo:2009ix}. It has been suggested that the weak coupling between gravity and matter might  suffice to suppress the percolation \cite{Pospelov:2010mp} but it is not clear how well this works in practice for Ho\v rava gravity \cite{Colombo:2014lta,Colombo:2015yha,Coates:2016zvg}.

Irrespective of the details, it is intuitive that experiments will impose a lower bound on $M_\star$. Interestingly, theoretical considerations can yield an upper bound. This is because the IR part of Ho\v rava gravity, or equivalently action \eqref{eq:khronometric} exhibits strong coupling at a certain scale $M_{SC}$ \cite{Papazoglou:2009fj}. In particular,  derivative self-interactions of the scalar mode compromise perturbativity. Provided that $M_{SC}$ is sufficiently high, strong coupling is not an issue for infrared viability. However, power-counting renormalizability for Ho\v rava gravity has been argued on the basis of perturbativity and hence, strong coupling is a threat to the original motivation of the theory  \cite{Papazoglou:2009fj}. It turns out that having $M_*<M_{SC}$ --- {\em i.e.}~having the new physics coming from the UV completion kick in at low enough energies --- can resolve the strong coupling problem \cite{Blas:2010hb}. However, as we will review in a bit more detail below, $M_{SC}$ is controlled by the couplings of the infrared part of the action, $(\lambda, \xi, \eta)$ or $(\alpha, \beta, \gamma)$, which satisfy the bounds given in the previous section. Hence, $M_\star$ ends up having to satisfy an upper bound.

The $M_{SC}$ can been calculated in the decoupling limit, which for action \eqref{eq:khronometric} corresponds to $M_{\AE}^2 \rightarrow \infty $ while keeping $\alpha M_{\AE}^2, \beta M_{\AE}^2, \gamma M_{\AE}^2$ fixed. This requires $\alpha, \beta , \gamma \to 0$.
%
A detailed calculation was presented in Ref.~\cite{Kimpton:2010xi}.  It is worth pointing out that $\beta$ has been set to zero there.  This can indeed be done without loss of generality by a suitable time rescaling in action \eqref{eq:fullaction} or field redefinitions in action \eqref{eq:khronometric}.  We have avoided it here because the rescaling affects the various speeds. We discuss in more details how what we report  corresponds to the results of Ref.~\cite{Kimpton:2010xi} in Appendix \ref{sec:Rescaling}. 

In the limit where $(\alpha, \beta, \gamma)\ll 1$ (consistent with the decoupling approximation), $M_{SC}$ can have one of two different values depending on the magnitude of $c_S^2$, namely%
\begin{equation}
\label{scscale}
M_{SC} \approx \Bigg\{\begin{array}{ll} 
    \sqrt{\alpha}M_{\AE} c_S^{3/2}  &\,, c_S^2<1 \\
    \sqrt{\alpha}M_{\AE} c_S^{-1/2} &\,, c_S^2>1 
   \end{array}.
\end{equation}
It is worth stressing that when $(\alpha, \beta, \gamma)\ll1$, and hence in the decoupling limit as well, $M_{\AE}\sim M_p$ and both scales could be taken to be equal to the gravitational coupling scale as measured by experiments.

\section{Allowed region of the parameter space} 

Prior to GW170817, ppN constraints were considered particularly restrictive because, assuming that $\beta\ll1$, they require $|\alpha-2\beta|\lesssim10^{-4}$. Hence, they restrict the 3-dimensional parameter space to a 2-dimensional surface with a width of $10^{-4}$. All other constraints were either one-sided or weaker, so it was common practice to impose $\alpha=2\beta$ in order to satisfy both ppN bounds to infinite accuracy (modulo tunings) and present graphically existing and new constraints on the $\alpha=2\beta$ plane, see for example Refs.~\cite{Barausse:2011pu,Barausse:2013nwa,Yagi:2013ava}. The double-sided constraint on $c_T$ or $\beta$ from GW170817 changes the picture drastically. The allowed region of the $\alpha=2\beta$ plane shrinks to a line segment with width less than $10^{-15}$, as one can always think of this plane as parametrised by $\beta$ itself and some other combination of the couplings ({\em c.f.}~with the figures in Refs.~\cite{Barausse:2011pu,Barausse:2013nwa,Yagi:2013ava}). 

Indeed it is now much more informative to focus on the $\beta=0$ plane of the parameter space or, more generally, $|\beta|\sim 10^{-15}$ sections, as in Figs.~\ref{fig:parameter_space_zero}-\ref{fig:parameter_space_negative}. The top panel of each figure shows the whole parameter space. The blue curve corresponds to $M_{SC}=1$ meV. Given that $M_*<M_{SC}$, the constraint $M_*>1$ meV, which is derived from binary black hole mergers \cite{Yunes:2016jcc}, excludes the region of parameter space corresponding to $M_{SC}<1$ meV. The red and brown curves correspond to $M_{SC}=1$ TeV and $M_{SC}=10^{10}$ GeV, respectively. These curves demonstrate how stricter bounds on $M_*$ would constraint the parameter space. The lower panels zoom on the region $\alpha,\gamma>10^{-20}$. The dotted, dashed and solid lines  correspond to choice of $\alpha$ and $\gamma$ that lead to $c_S=1$, $c_S=10$ and $c_S=1000$, respectively. They have been included to highlight that the $c_S$ remains virtually unconstrained. It should be stressed that we have not imposed the vacuum Cherenkov constraints. Imposing them in a conservative fashion corresponds to excluding the part of the shaded region above the  $c_S^2=1$ dotted line in each plot.

\section{Discussion}

The detection of gravitational waves with an electromagnetic counterpart (GW170817), emitted by a binary neutron star merger, has put a stringent constraint on the speed of tensor modes. This translates to the tightest constraint so far in one of the parameters of Ho\v{r}ava gravity and it motivates revisiting the allowed region of the 3-dimensional parameter space. Instead of focussing on the 2-d plane that leads to exact agreement with general relativity in the weak field limit, as had been done in the literature so far, we focussed on  2-dimensional sections that satisfy the gravitational wave speed constraint and considered all other known constraints. The graphic representation of these constraints turns out to be quite illuminating in at least two respects:
\begin{enumerate}
\item The strong coupling scale $M_{SC}$ is bounded from below by the Lorentz breaking scale $M_\star$, as discussed extensively in the text. $M_\star$ is in turn bounded from below by observations that probe the higher order terms in the dispersion relation. Improving bounds on $M_\star$ can hence reduce the parameter space significantly or rule out Ho\v rava gravity entirely as a perturbatively renormalizable theory of gravity. 
\item Even though the constraints on the parameters $\alpha$, $\beta$ and $\gamma$ are very tight, the speed of the scalar polarisation remains virtually unconstrained. This stems from the type of dependence $c_S$ has on these parameters and, more fundamentally, from the fact that the limit to general relativity is not smooth. Remarkably, even a very mild constraint on $c_S$ would rule out a vast portion of the parameter space. 
\end{enumerate}
The aboves motivate finding novel ways to improve constraints on $M_\star$ and $c_S$. 

The prospects of measuring or constraining $c_S$ and its importance in the context of Lorentz violations in gravity have been recently discussed in Ref.~\cite{Sotiriou:2017obf}. In the context of Ho\v rava gravity, one could hope to obtain stricter bounds  on $M_\star$ and $c_S$ by precise gravitational waveform modelling. In principle, trustworthy bounds on $M_\star$ could also be obtained from nongravitational experiments (along the lines of Ref.~\cite{Liberati:2012jf}), if percolations of Lorentz symmetry from gravity to matter were well understood. Both prospects are quite challenging but would significantly enhance our understanding of the role of Lorentz symmetry in gravitation. 


\acknowledgments{We are grateful to Andrew Coates, Ted Jacobson and Tony Padilla for illuminating discussions. AEG acknowledges financial support from the European Research Council (ERC) under the European Union's Horizon 2020
research and innovation programme (grant agreement 646702 ”CosTesGrav”). M.S. is supported by the Royal Commission for the Exhibition of 1851. TPS acknowledges funding from the European
Research Council under the European Union's Seventh Framework Programme
(FP7/2007-2013) / ERC grant agreement n.~306425 ``Challenging General
Relativity'' and partial support from the STFC Consolidated Grant ST/P000703/1. TPS would also like to acknowledge networking support by the COST Action GWverse
CA16104.}

\appendix
\section{Correspondence with hypersurface orthogonal Einstein-\AE{}ther theory}
The action for Einstein-\AE{}ther theory 
\cite{Jacobson:2000xp}
is given by 
\begin{align}
S=\frac{M^2_\textrm{\AE{}}}{2}\int \sqrt{-g} \,d^4x [& ~^{(4)}R-c_1 \nabla_\mu u_\nu \nabla^\mu u^\nu
\nonumber\\
&-c_2(\nabla_\mu u^\mu)^2-c_3 \nabla_\mu u_\nu\nabla^\nu u^\mu
\nonumber\\
&+c_4 u_\alpha u_\beta \nabla^\alpha u_\mu\,\nabla^\beta u^\mu]\,,
\end{align}
where the norm of the \ae{}ther field is constrained by $u^\mu u_\mu =-1$. If the \ae{}ther field is hypersurface orthogonal it can simply be written as the (normalized) gradient of a single scalar:
\begin{equation} 
u_\mu = - \frac{\nabla_\mu \phi}{\sqrt{-\nabla_\nu\phi \nabla^\nu\phi}}\,.
\end{equation}
When this form of the vector field is imposed at the level of the action, {\em i.e.} before the variation, the $c_1$ term is no longer independent, and can be written as a combination of the $c_3$ and $c_4$. The resulting action matches the one in Eq.~\eqref{eq:khronometric}, and the correspondence of parameters is
\begin{equation}
\alpha= c_1+c_4\,,\;
\beta = c_1+c_3\,,\;
\gamma=c_2\,.
\end{equation}
\label{sec:aetherapp}

\section{Time rescaling and the value of $\beta$ parameter}
\label{sec:Rescaling}

 In action \eqref{eq:fullaction} one can set $\xi=1$ by performing the time rescaling 
\beq\label{time_rescaling}
d\tilde t=\sqrt{\xi}dt\,.
\eeq
Given the correspondence of couplings in eq.~\eqref{corres}, in the covariant picture of action \eqref{eq:khronometric} this maps to $\beta=0$. 
Note that $\xi>0$ (or $\beta<1$) is required for stability, see Sec. \ref{section_constraint}.
Time rescalings do not leave speeds invariant clearly and the specific one corresponds to choosing the time coordinate such that the speed of tensor gravitational waves is set to 1. This becomes clearer if one tries to set $\beta=0$ directly in the covariant setup of action \eqref{eq:khronometric}, as it requires a particular metric redefinition and a $u_\mu$ rescaling of the type discussed in Ref.~\cite{Foster:2005ec}. Such redefinitions leave the action formally invariant and for the specific ones that leads to $\beta=0$ the new metric defines the null propagation cones  of spin-2 gravitons. It should be emphasized that 
the speed of light also changes and, assuming it was 1 initially, it becomes $\sqrt{1-\beta}$ after either of the two equivalent procedures discussed above. 
The complete mapping of couplings is 
\bea
&&\tilde \alpha=\alpha\, ,\qquad \tilde \gamma=\frac{\gamma+\beta}{1-\beta}\notag\\
&&\tilde \beta=0\, ,\qquad  \tilde M_{\AE}^2= \sqrt{1-\beta} M_{AE}^2,
\eea
where an overtilde denotes the new couplings. 

Ref.\cite{Kimpton:2010xi} performs calculations in the decoupling limit but it resorts to this limit after having set $\tilde{\beta}$ to zero as described above. Hence, the limit is taken to be $\tilde{M}_{\AE}^2 \rightarrow \infty $ while keeping $\tilde{\alpha} \tilde{M}_{\AE}^2, \tilde{\gamma} \tilde{M}_{\AE}^2$ fixed. This implies that $\tilde{\alpha}, \tilde{\gamma} \to 0$.
while there is no further explicit reference to $\beta$.
The speed of the scalar polarisation  in this limit is $\tilde{c}_{\rm dec}^2=\tilde \gamma/\tilde \alpha$.
In the process of determining the smallest  suppression scale for cubic interactions, and therefore identifying the strong coupling scale, the authors of Ref.\cite{Kimpton:2010xi} impose $\tilde c_{\rm dec} \leq 1$, with the justification that it is preferable to avoid superluminality. Under this assumption the strong coupling scale turns out to be  
\begin{equation}
\tilde M_{SC}=\sqrt{\tilde \alpha}\tilde M_{\AE} \tilde c_{\rm dec}^{3/2}\,, 
\label{strong_coupling}
\end{equation}
as this scale  has the largest positive exponent for $\tilde c_{\rm dec}$.
%
A minor points of caution is that, after the time rescaling \eqref{time_rescaling} the speed of light is no longer unity but $\sqrt{1-\beta}$. Hence, superluminal propagation in the decoupling limit corresponds to  $\tilde c_{\rm dec}^2>1-\beta$. More importantly, in a Lorentz violating theory with a preferred foliation there is no pertinent reason to exclude superluminal propagation. On the contrary, the vacuum Cherenkov bounds disfavour subluminal modes.

The decoupling limit as defined in Sec.~\ref{sec:dec} is $M_{\AE}^2 \rightarrow \infty $ while keeping $\alpha M_{\AE}^2, \beta M_{\AE}^2, \gamma M_{\AE}^2$ fixed. This requires $\alpha, \beta , \gamma \to 0$, so it does not correspond exactly to the limit taken in Ref.\cite{Kimpton:2010xi}. However, when $\beta\to 0$ there is perfect  agreement. Moreover, $\tilde c_{\rm dec}^2\sim c^2_S$ and the first point of caution above becomes moot. Indeed, 
following the line of Ref.~\cite{Kimpton:2010xi}  but relaxing the assumption $\tilde c_{\rm dec}\leq 1$, the strong coupling scale is
\bea
\!\!\! M_{SC}=\Bigg\{\begin{array}{ll} 
   \! \sqrt{\alpha} M_{\AE} c_S^{3/2} (1-\beta)^{3/4} \,,\!\!& c_S^2 (1-\beta)<1\\
    \!\sqrt{\alpha}M_{\AE} c_S^{-1/2}(1-\beta)^{-1/4} \,,\!\!& c_S^2 (1-\beta)>1\nonumber
   \end{array}
\eea
where we have purposefully not taken the limit $\beta\to 0$. 
These expressions agree with that of  eq.~\eqref{scscale} when $\alpha,\beta,\gamma\ll 1$.


\begin{thebibliography}{99}

\bibitem{Horava:2009uw} 
  P.~Horava,
  Phys.\ Rev.\ D {\bf 79}, 084008 (2009)
  doi:10.1103/PhysRevD.79.084008
  [arXiv:0901.3775 [hep-th]].
  

\bibitem{Blas:2009qj} 
  D.~Blas, O.~Pujolas and S.~Sibiryakov,
  Phys.\ Rev.\ Lett.\  {\bf 104}, 181302 (2010)
  doi:10.1103/PhysRevLett.104.181302
  [arXiv:0909.3525 [hep-th]].
  
\bibitem{Blas:2010hb} 
  D.~Blas, O.~Pujolas and S.~Sibiryakov,
  JHEP {\bf 1104}, 018 (2011)
  doi:10.1007/JHEP04(2011)018
  [arXiv:1007.3503 [hep-th]].

  
\bibitem{Jacobson:2000xp} 
  T.~Jacobson and D.~Mattingly,
  Phys.\ Rev.\ D {\bf 64}, 024028 (2001)
  doi:10.1103/PhysRevD.64.024028
  [gr-qc/0007031].
  
\bibitem{Jacobson:2010mx} 
  T.~Jacobson,
  Phys.\ Rev.\ D {\bf 81}, 101502 (2010)
  Erratum: [Phys.\ Rev.\ D {\bf 82}, 129901 (2010)]
  doi:10.1103/PhysRevD.82.129901, 10.1103/PhysRevD.81.101502
  [arXiv:1001.4823 [hep-th]].
  
\bibitem{Monitor:2017mdv} 
  B.~P.~Abbott {\it et al.} [LIGO Scientific and Virgo and Fermi-GBM and INTEGRAL Collaborations],
  Astrophys.\ J.\  {\bf 848}, no. 2, L13 (2017)
  doi:10.3847/2041-8213/aa920c
  [arXiv:1710.05834 [astro-ph.HE]].
  
\bibitem{Sotiriou:2017obf} 
  T.~P.~Sotiriou,
  arXiv:1709.00940 [gr-qc].
  
\bibitem{Sotiriou:2009gy} 
  T.~P.~Sotiriou, M.~Visser and S.~Weinfurtner,
  Phys.\ Rev.\ Lett.\  {\bf 102}, 251601 (2009)
  doi:10.1103/PhysRevLett.102.251601
  [arXiv:0904.4464 [hep-th]].
  
  
\bibitem{Weinfurtner:2010hz} 
  S.~Weinfurtner, T.~P.~Sotiriou and M.~Visser,
  J.\ Phys.\ Conf.\ Ser.\  {\bf 222}, 012054 (2010)
  doi:10.1088/1742-6596/222/1/012054
  [arXiv:1002.0308 [gr-qc]].
  
\bibitem{Horava:2010zj} 
  P.~Horava and C.~M.~Melby-Thompson,
  Phys.\ Rev.\ D {\bf 82}, 064027 (2010)
  doi:10.1103/PhysRevD.82.064027
  [arXiv:1007.2410 [hep-th]].
  
\bibitem{Vernieri:2011aa} 
  D.~Vernieri and T.~P.~Sotiriou,
  Phys.\ Rev.\ D {\bf 85}, 064003 (2012)
  doi:10.1103/PhysRevD.85.069901, 10.1103/PhysRevD.85.064003
  [arXiv:1112.3385 [hep-th]].
  
\bibitem{Vernieri:2012ms} 
  D.~Vernieri and T.~P.~Sotiriou,
  J.\ Phys.\ Conf.\ Ser.\  {\bf 453}, 012022 (2013)
  doi:10.1088/1742-6596/453/1/012022
  [arXiv:1212.4402 [hep-th]].
  
  
\bibitem{Barvinsky:2015kil} 
  A.~O.~Barvinsky, D.~Blas, M.~Herrero-Valea, S.~M.~Sibiryakov and C.~F.~Steinwachs,
  Phys.\ Rev.\ D {\bf 93}, no. 6, 064022 (2016)
  doi:10.1103/PhysRevD.93.064022
  [arXiv:1512.02250 [hep-th]].
  
\bibitem{Sotiriou:2011dr} 
  T.~P.~Sotiriou, M.~Visser and S.~Weinfurtner,
  Phys.\ Rev.\ D {\bf 83}, 124021 (2011)
  doi:10.1103/PhysRevD.83.124021
  [arXiv:1103.3013 [hep-th]].
  
\bibitem{Barvinsky:2017kob} 
  A.~O.~Barvinsky, D.~Blas, M.~Herrero-Valea, S.~M.~Sibiryakov and C.~F.~Steinwachs,
  arXiv:1706.06809 [hep-th].
  
\bibitem{Sotiriou:2009bx} 
  T.~P.~Sotiriou, M.~Visser and S.~Weinfurtner,
  JHEP {\bf 0910}, 033 (2009)
  doi:10.1088/1126-6708/2009/10/033
  [arXiv:0905.2798 [hep-th]].
  
\bibitem{Charmousis:2009tc} 
  C.~Charmousis, G.~Niz, A.~Padilla and P.~M.~Saffin,
  JHEP {\bf 0908}, 070 (2009)
  doi:10.1088/1126-6708/2009/08/070
  [arXiv:0905.2579 [hep-th]].
  
\bibitem{Blas:2009yd} 
  D.~Blas, O.~Pujolas and S.~Sibiryakov,
  JHEP {\bf 0910}, 029 (2009)
  doi:10.1088/1126-6708/2009/10/029
  [arXiv:0906.3046 [hep-th]].
  
\bibitem{Koyama:2009hc} 
  K.~Koyama and F.~Arroja,
  JHEP {\bf 1003}, 061 (2010)
  doi:10.1007/JHEP03(2010)061
  [arXiv:0910.1998 [hep-th]].
  
\bibitem{Papazoglou:2009fj} 
  A.~Papazoglou and T.~P.~Sotiriou,
  Phys.\ Lett.\ B {\bf 685}, 197 (2010)
  doi:10.1016/j.physletb.2010.01.054
  [arXiv:0911.1299 [hep-th]].


  
\bibitem{Kimpton:2010xi} 
  I.~Kimpton and A.~Padilla,
  JHEP {\bf 1007}, 014 (2010)
  doi:10.1007/JHEP07(2010)014
  [arXiv:1003.5666 [hep-th]].
  
\bibitem{Mukohyama:2010xz} 
  S.~Mukohyama,
  Class.\ Quant.\ Grav.\  {\bf 27}, 223101 (2010)
  doi:10.1088/0264-9381/27/22/223101
  [arXiv:1007.5199 [hep-th]].
  
\bibitem{Jacobson:2008aj} 
  T.~Jacobson,
  PoS QG {\bf -PH}, 020 (2007)
  [arXiv:0801.1547 [gr-qc]].

\bibitem{Chen:2000xxa} 
  X.~l.~Chen, R.~J.~Scherrer and G.~Steigman,
  Phys.\ Rev.\ D {\bf 63}, 123504 (2001)
  doi:10.1103/PhysRevD.63.123504
  [astro-ph/0011531].
  
\bibitem{Carroll:2004ai} 
  S.~M.~Carroll and E.~A.~Lim,
  Phys.\ Rev.\ D {\bf 70}, 123525 (2004)
  doi:10.1103/PhysRevD.70.123525
  [hep-th/0407149].
  

\bibitem{Izotov:2014fga}
  Y.~I.~Izotov, T.~X.~Thuan and N.~G.~Guseva,
  Mon.\ Not.\ Roy.\ Astron.\ Soc.\  {\bf 445} (2014) no.1,  778
  doi:10.1093/mnras/stu1771
  [arXiv:1408.6953 [astro-ph.CO]].
  
\bibitem{Aver:2015iza} 
  E.~Aver, K.~A.~Olive and E.~D.~Skillman,
  JCAP {\bf 1507}, no. 07, 011 (2015)
  doi:10.1088/1475-7516/2015/07/011
  [arXiv:1503.08146 [astro-ph.CO]].
  
\bibitem{Patrignani:2016xqp} 
  C.~Patrignani {\it et al.} [Particle Data Group],
  Chin.\ Phys.\ C {\bf 40}, no. 10, 100001 (2016).
  doi:10.1088/1674-1137/40/10/100001
  
 
\bibitem{Moore:2001bv} 
  G.~D.~Moore and A.~E.~Nelson,
  JHEP {\bf 0109}, 023 (2001)
  doi:10.1088/1126-6708/2001/09/023
  [hep-ph/0106220].

\bibitem{Elliott:2005va} 
  J.~W.~Elliott, G.~D.~Moore and H.~Stoica,
  JHEP {\bf 0508}, 066 (2005)
  doi:10.1088/1126-6708/2005/08/066
  [hep-ph/0505211].
  
  \bibitem{Kiyota:2015dla}
  S.~Kiyota and K.~Yamamoto,
  Phys.\ Rev.\ D {\bf 92} (2015) no.10,  104036
  doi:10.1103/PhysRevD.92.104036
  [arXiv:1509.00610 [gr-qc]].
  
\bibitem{Yunes:2016jcc}
N.~Yunes, K.~Yagi and F.~Pretorius,
Phys.\ Rev.\ D {\bf94}, 084002 (2016)
doi:10.1103/PhysRevD.94.084002
[Arxiv:1603.08955[gr-qc]],

\bibitem{Will:2005va} 
  C.~M.~Will,
  Living Rev.\ Rel.\  {\bf 9}, 3 (2006)
  doi:10.12942/lrr-2006-3
  [gr-qc/0510072].

\bibitem{Blas:2011zd} 
  D.~Blas and H.~Sanctuary,
  Phys.\ Rev.\ D {\bf 84}, 064004 (2011)
  doi:10.1103/PhysRevD.84.064004
  [arXiv:1105.5149 [gr-qc]].
  
  
\bibitem{Yagi:2013ava} 
  K.~Yagi, D.~Blas, E.~Barausse and N.~Yunes,
  Phys.\ Rev.\ D {\bf 89}, no. 8, 084067 (2014)
  Erratum: [Phys.\ Rev.\ D {\bf 90}, no. 6, 069902 (2014)]
  Erratum: [Phys.\ Rev.\ D {\bf 90}, no. 6, 069901 (2014)]
  doi:10.1103/PhysRevD.90.069902, 10.1103/PhysRevD.90.069901, 10.1103/PhysRevD.89.084067
  [arXiv:1311.7144 [gr-qc]].
  
\bibitem{Barausse:2011pu} 
  E.~Barausse, T.~Jacobson and T.~P.~Sotiriou,
  Phys.\ Rev.\ D {\bf 83}, 124043 (2011)
  doi:10.1103/PhysRevD.83.124043
  [arXiv:1104.2889 [gr-qc]].
  
\bibitem{Barausse:2012ny} 
  E.~Barausse and T.~P.~Sotiriou,
  Phys.\ Rev.\ Lett.\  {\bf 109}, 181101 (2012)
  Erratum: [Phys.\ Rev.\ Lett.\  {\bf 110}, no. 3, 039902 (2013)]
  doi:10.1103/PhysRevLett.110.039902, 10.1103/PhysRevLett.109.181101
  [arXiv:1207.6370 [gr-qc]].
  
\bibitem{Barausse:2012qh} 
  E.~Barausse and T.~P.~Sotiriou,
  Phys.\ Rev.\ D {\bf 87}, 087504 (2013)
  doi:10.1103/PhysRevD.87.087504
  [arXiv:1212.1334 [gr-qc]].
  
\bibitem{Barausse:2013nwa} 
  E.~Barausse and T.~P.~Sotiriou,
  Class.\ Quant.\ Grav.\  {\bf 30}, 244010 (2013)
  doi:10.1088/0264-9381/30/24/244010
  [arXiv:1307.3359 [gr-qc]].

\bibitem{Abbott:2017vtc} 
  B.~P.~Abbott {\it et al.} [LIGO Scientific and VIRGO Collaborations],
  Phys.\ Rev.\ Lett.\  {\bf 118}, no. 22, 221101 (2017)
  doi:10.1103/PhysRevLett.118.221101
  [arXiv:1706.01812 [gr-qc]].

  
\bibitem{Adelberger:2009zz} 
  E.~G.~Adelberger, J.~H.~Gundlach, B.~R.~Heckel, S.~Hoedl and S.~Schlamminger,
  Prog.\ Part.\ Nucl.\ Phys.\  {\bf 62}, 102 (2009).
  doi:10.1016/j.ppnp.2008.08.002



\bibitem{Liberati:2012jf} 
  S.~Liberati, L.~Maccione and T.~P.~Sotiriou,
  Phys.\ Rev.\ Lett.\  {\bf 109}, 151602 (2012)
  doi:10.1103/PhysRevLett.109.151602
  [arXiv:1207.0670 [gr-qc]].

\bibitem{Iengo:2009ix} 
  R.~Iengo, J.~G.~Russo and M.~Serone,
  JHEP {\bf 0911}, 020 (2009)
  doi:10.1088/1126-6708/2009/11/020
  [arXiv:0906.3477 [hep-th]].
  
\bibitem{Pospelov:2010mp} 
  M.~Pospelov and Y.~Shang,
  Phys.\ Rev.\ D {\bf 85}, 105001 (2012)
  doi:10.1103/PhysRevD.85.105001
  [arXiv:1010.5249 [hep-th]].
  
\bibitem{Colombo:2014lta} 
  M.~Colombo, A.~E.~Gumrukcuoglu and T.~P.~Sotiriou,
  Phys.\ Rev.\ D {\bf 91}, no. 4, 044021 (2015)
  doi:10.1103/PhysRevD.91.044021
  [arXiv:1410.6360 [hep-th]].
  
\bibitem{Colombo:2015yha} 
  M.~Colombo, A.~E.~Gumrukcuoglu and T.~P.~Sotiriou,
  Phys.\ Rev.\ D {\bf 92}, no. 6, 064037 (2015)
  doi:10.1103/PhysRevD.92.064037
  [arXiv:1503.07544 [hep-th]].

  
\bibitem{Coates:2016zvg} 
  A.~Coates, M.~Colombo, A.~E.~Gumrukcuoglu and T.~P.~Sotiriou,
  Phys.\ Rev.\ D {\bf 94}, no. 8, 084014 (2016)
  doi:10.1103/PhysRevD.94.084014
  [arXiv:1604.04215 [hep-th]].
  
  



  

\bibitem{Foster:2005ec} 
  B.~Z.~Foster,
  Phys.\ Rev.\ D {\bf 72}, 044017 (2005)
  doi:10.1103/PhysRevD.72.044017
  [gr-qc/0502066].
  
  \end{thebibliography}
\end{document}